\documentclass[preprint,12pt, a4paper]{elsarticle}

\usepackage{amssymb}
\usepackage{hyperref}
\usepackage{booktabs}
\usepackage{enumitem}
\usepackage{caption} 
\usepackage{listings}
\usepackage{amsfonts}
\usepackage{bibentry}
\usepackage{amsmath}
\nobibliography*
\usepackage{tikz}
\usetikzlibrary{arrows.meta, positioning, shapes, fit, calc}

\definecolor{geeBlue}{RGB}{47,85,151}
\definecolor{kronPurple}{RGB}{128,0,128}
\definecolor{espOrange}{RGB}{227,114,34}
\definecolor{carryGreen}{RGB}{0,150,136}

\tikzset{
  font=\small,
  startstop/.style={
    rectangle, rounded corners, draw=black, ultra thick,
    fill=gray!10, minimum width=4.6cm, minimum height=0.9cm, align=center
  },
  process/.style={
    rectangle, rounded corners, draw=carryGreen!50!black, thick,
    fill=carryGreen!10, text width=7.5cm, minimum height=1.2cm, align=left, inner sep=6pt
  },
  processGEE/.style={
    rectangle, rounded corners, draw=geeBlue!60!black, thick,
    fill=geeBlue!8, text width=5.0cm, minimum height=1.2cm, align=left, inner sep=6pt
  },
  processKron/.style={
    rectangle, rounded corners, draw=kronPurple!80!black, thick,
    fill=kronPurple!10, text width=4.0cm, minimum height=1.2cm, align=left, inner sep=6pt
  },
  processESP/.style={
    rectangle, rounded corners, draw=espOrange!80!black, thick,
    fill=espOrange!10, text width=4.0cm, minimum height=1.2cm, align=left, inner sep=6pt
  },
  arrow/.style={-{Latex[length=3mm,width=2mm]}, very thick},
  note/.style={rectangle, draw=gray!60, fill=white, rounded corners,
               align=left, inner sep=4pt, font=\footnotesize, text width=10.8cm},
  legendbox/.style={rectangle, draw=black, rounded corners, fill=white, inner sep=6pt}
}

\usepackage{pifont}
\newcommand{\cmark}{\ding{51}}  
\newcommand{\xmark}{\ding{55}}  
\newcommand{\hmark}{\ding{182}} 
\journal{SoftwareX}

\begin{document}
\renewcommand{\labelenumii}{\arabic{enumi}.\arabic{enumii}}

\begin{frontmatter}



\title{\texttt{CrossCarry}: An \texttt{R} package for the  analysis of data from a crossover design with GEE}


\author[label1,label2,label3,label4]{N.A. Cruz\corref{cor1}}
\author[label5]{O.O. Melo}
\author[label6]{C.A. Martinez}
\author[label1,label2,label3,label4]{R. Alberich}
\cortext[cor1]{Universitat de les Illes Balears, Departament de Matemàtiques i Informàtica, Phone: +34 637 54 6888, Palma de Mallorca, España, nelson-alirio.cruz@uib.es}
\address[label1]{Artificial Intelligence Research Institute of the Balearic Islands (IAIB), Department of Mathematics and Computer Science, University of the Balearic Islands, Palma 07122, Spain}
\address[label2]{Health Research Institute of the Balearic Islands (IdISBa), Palma 07010, Spain}
\address[label3]{Laboratory of Artificial Intelligence Applications (LAIA@UIB), Department of Mathematics and Computer Science, University of the Balearic Islands, Palma 07122, Spain}
\address[label4]{Data Modelling and Statistical Learning (MoDAE), Department of Mathematics and Computer Science, University of the Balearic Islands, Palma 07122, Spain}
\address[label5]{Departamento de Estadística, Facultad de Ciencias,  Universidad Nacional de Colombia, Bogota, Colombia}
\address[label6]{Departamento de Producción Animal, Facultad de Medicina Veterinaria y Zootecnia,  Universidad Nacional de Colombia, Bogota, Colombia}
\begin{abstract}
\textit{Crossover designs are widely applied in medicine, agriculture, and other biological sciences, yet their analysis remains challenging due to longitudinal observations within each unit and the presence of carry-over effects. Despite their prevalence, there is no comprehensive \texttt{R} package dedicated to the statistical modeling of crossover data. The \texttt{CrossCarry} package addresses this gap by providing a flexible and open-source framework for analyzing any crossover design with response variables from the exponential family, with or without washout periods. It extends the generalized estimating equations (GEE) methodology by incorporating correlation structures specifically tailored to crossover data, capturing both within- and between-period dependencies. Moreover, \texttt{CrossCarry} integrates a parametric component for treatment effects and a nonparametric spline-based component for time and carry-over effects. This combination allows users to model complex correlation patterns and temporal structures with minimal coding effort. By offering a domain-independent implementation of advanced statistical methodology, \texttt{CrossCarry} facilitates reproducible research and promotes the reuse of robust analytical tools across disciplines. Its potential applications span medical trials, agricultural field experiments, and other areas where crossover designs are essential, thus contributing to broader scientific discovery and cross-domain methodological standardization.
}

\end{abstract}


\end{frontmatter}

\section*{Metadata}

\begin{table}[!h]
\begin{tabular}{|l|p{6.5cm}|p{6.5cm}|}
\hline
\textbf{Nr.} & \textbf{Code metadata description} & \textbf{Metadata} \\
\hline
C1 & Current code version & R Version - 1.0.0, \\
\hline
C2 & Permanent link to code/repository used for this code version &  {\url{https://doi.org/10.5281/zenodo.17611428}} \\
\hline
C3  & Permanent link to Reproducible Capsule &  \url{https://github.com/Cruzalirio/CrossCarry/tree/master/examples}\\
\hline
C4 & Legal Code License   & GNU General Public License v3.0 \\
\hline
C5 & Code versioning system used & git \\
\hline
C6 & Software code languages, tools, and services used & R \\
\hline
C7 & Compilation requirements, operating environments \& dependencies & Depends are declared in \url{https://cran.r-project.org/web/packages/CrossCarry/index.html} \\
\hline
C8 & If available Link to developer documentation/manual & \url{https://cran.r-project.org/web/packages/CrossCarry/CrossCarry.pdf} \\
\hline
C9 & Support email for questions & \url{nelson-alirio.cruz@uib.es}\\
\hline
\end{tabular}
\caption{Code metadata}
\label{codeMetadata} 
\end{table}

\section{Motivation and significance}
Crossover designs are a cornerstone of experimental methodology in medicine, agriculture, and the biological sciences, as they allow comparisons of treatments within the same experimental units by administering sequences of treatments across multiple periods \citep{ken15}. By construction, these designs generate longitudinal data and often induce carry-over effects, where responses in later periods are influenced by treatments received earlier. Correctly accounting for such dependencies, alongside the complex within- and between-period correlations, is crucial for valid inference. More generally, when units are observed sequentially, the design is categorized as longitudinal or repeated measures \citep{hk}; crossover designs introduce an additional layer of complexity due to their sequential treatment structure.

From a methodological standpoint, generalized linear mixed models (GLMM) and generalized estimating equations (GEE) are the dominant approaches for handling responses from the exponential family. GLMMs model random effects under explicit distributional assumptions, whereas GEEs provide robust inference for marginal treatment effects using sandwich estimators \citep{liang1986longitudinal, basu2010joint, zhang2012new}. The latter are particularly attractive when the focus is on treatment comparisons rather than random effects. However, correlation structures available in popular R packages such as \texttt{geepack} and \texttt{geeM} (exchangeable, autoregressive, or unstructured) do not capture the dependencies specific to crossover designs, especially in the presence of repeated measurements within periods and complex carry-over effects \citep{hojsgaard2006r, geeM}.

{GEE have also become a standard tool in a wide range of applied fields (including biomedical, agricultural, and biological sciences) due to their robustness and flexibility for modeling correlated outcomes \citep{GEE1}. Recent studies highlight the versatility of GEE-based models in settings involving patient-level clinical trajectories, omics-derived biomarkers, and high-dimensional longitudinal measurements, where inference on marginal effects remains the primary target \citep{GEE2, GEE3}. These applications routinely involve challenges such as within-subject correlation, irregular follow-up times, and missing data \citep{GEE4}}

{Crossover trials share many of these challenges and pose additional complications related to period and carry-over effects.} Recent methodological work has addressed these shortcomings. \cite{cruz2023analysis} unified the use of GEE for crossover designs with a single measurement per period under classical correlation structures. When repeated measurements are recorded within each period, these structures become inadequate; to overcome this, \cite{cruz2023correlation} proposed Kronecker-product correlation structures that jointly model within- and between-period dependence. For carry-over effects, \cite{cruz2023semiparametric} developed a semiparametric framework that employs splines within the estimating equations, enabling the estimation of both simple and complex forms of carry-over. Together with critical discussions on identifiability and limitations of classical carry-over models \citep{senn1992simple}, these advances provide the theoretical foundation for software implementation.

{ In preparing Table~\ref{tab:comparison}, we conducted a review of the R ecosystem—searching CRAN, Bioconductor, and actively developed GitHub repositories—to identify software relevant to crossover designs. This exploration confirmed that, despite notable progress, there is still no comprehensive and actively maintained package that integrates recent methodological developments for analyzing crossover data. Existing tools provide only partial support in the precise sense that each package addresses a specific stage of the workflow—such as design generation, power calculation, or randomization—while omitting the modelling capabilities required for real-data analysis.} For instance, \texttt{clusterPower} focuses exclusively on power computation for cluster and cluster-randomized crossover trials \citep{paquetepower}; \texttt{crossdes} generates balanced designs but assumes negligible carry-over effects and does not implement estimation procedures \citep{crossdess}; \texttt{randomizeBE} provides subject-level randomization tools for bioequivalence settings \citep{randomizeBE}; and the \texttt{Crossover} package, formerly used for simple $2\times2$ designs under normality assumptions, lacks support for non-normal outcomes, flexible carry-over structures, and repeated measurements within periods. {As summarized in Table~\ref{tab:comparison}, these packages are therefore valuable for isolated tasks, yet none offer an integrated framework capable of accommodating non-normal responses, structured period and carry-over effects, and tailored working correlation matrices.}

The \texttt{CrossCarry} package was developed to fill this gap.  Unlike other tools, \texttt{CrossCarry} integrates in a single framework: (i) analyze designs with or without washout periods; (ii) accommodate responses from the exponential family; (iii) model treatment, time, and carry-over effects through parametric and nonparametric components; and (iv) handle repeated measurements within periods through tailored correlation structures. 
\begin{table}[ht]
\centering
\small
\begin{tabular}{|p{5.5cm}|p{0.75cm}|p{0.75cm}|p{0.75cm}|p{0.75cm}|p{0.75cm}|p{0.75cm}|p{0.75cm}|}
\hline
\textbf{Feature} &\textbf{CC} &\textbf{GP} &\textbf{GM} &\textbf{CO} &\textbf{CD} &\textbf{rBE} &\textbf{ CP} \\
\hline
GEE crossover end-to-end     & \cmark & \hmark & \hmark & \xmark & \xmark & \hmark & \xmark \\
\hline
Repeated measures within period & \cmark & \hmark & \hmark & \xmark & \xmark & \hmark & \xmark \\
\hline
Kronecker correlation        & \cmark & \xmark & \xmark & \xmark & \xmark & \xmark & \xmark \\
\hline
Spline carry-over/time       & \cmark & \xmark & \xmark & \xmark & \xmark & \xmark & \xmark \\
\hline
Visual diagnostics           & \cmark & \hmark & \hmark & \xmark & \xmark & \xmark & \xmark \\
\hline
Sample size/power            & \xmark & \xmark & \xmark & \xmark & \xmark & \xmark & \cmark \\
\hline
Randomization tools          & \xmark & \xmark & \xmark & \cmark & \cmark & \cmark & \xmark \\
\hline
Non-normal responses          & \cmark & \hmark & \hmark & \xmark & \xmark & \xmark & \xmark \\
\hline
\end{tabular}
\caption{Comparative features of \texttt{CrossCarry} and related R packages. 
Legend: \cmark available; \hmark partial; \xmark not available. Abbreviations: CC = \texttt{CrossCarry}, GP = \texttt{geepack}, GM = \texttt{geeM}, 
CO = \texttt{Crossover}, CD = \texttt{crossdes}, rBE = \texttt{randomizeBE}, 
CP = \texttt{clusterPower}.}
\label{tab:comparison}
\end{table}
It is important to clarify that the \texttt{CrossCarry} package is not intended 
to replace these other tools, but rather to complement them. Users may rely on 
packages such as \texttt{Crossover} or \texttt{crossdes} for the design stage, 
\texttt{clusterPower} for sample size, or general-purpose GEE/GLMM software for 
alternative modeling strategies. The contribution of \texttt{CrossCarry} 
is to provide a dedicated implementation of methods that have 
been theoretically validated and empirically tested in recent methodological 
papers \cite{cruz2023analysis,cruz2023correlation,cruz2023semiparametric}. 

In particular, those works already present simulation studies assessing bias, RMSE, and coverage under varying correlation structures, missing data, and non-Gaussian responses.
The relevance of this contribution is underscored by its broad applicability. Recent studies have demonstrated the importance of advanced crossover methodology in diverse domains: empirical software engineering {\citep{vegas2016crossover,kitchenham2021importance,frattini2024crossover,frattini2025good},} randomized neurology trials \citep{yu2025effect}, biomedical research using bioelectrical impedance \citep{knoll2025can}, Bayesian design for binary crossover trials \citep{singh2024bayesian}, clinical studies on cannabidiol and alcohol use disorder \citep{hurzeler2024neurobehavioural, hurzeler2025cannabidiol}, and educational research on digital storytelling and knowledge retention \citep{ginting2024effects}. By translating recent methodological advances into an open and reproducible software environment, \texttt{CrossCarry} enables researchers across disciplines to adopt robust statistical practices for crossover designs.

\section{Software Functionalities}

The \texttt{CrossCarry} package provides methods for the analysis of crossover experimental designs with repeated measures. The package is implemented in R and organizes its functionalities into four main functions: \texttt{createCarry}, \texttt{CrossGEE}, \texttt{CrossGEEKron}, and \texttt{CrossGEESP} as seen in Figure \ref{fig:CrossCarryFlowStyled}.

\begin{figure}[!ht]
\centering
\begin{tikzpicture}[node distance=1.7cm]

\node (start) [startstop] {\textbf{Input}: Crossover design data\\
\textbf{Required components}:
\texttt{treatment} (factor/character), \\
\texttt{period} (integer/numeric), \texttt{id} (factor/character),\\
\textbf{Optional components}:\texttt{time} (numeric), \texttt{covariates} (optional)};
\node (carry) [process, below=0.6cm of start]
{\textbf{\texttt{createCarry}} — carry-over dummies\\
\textbf{Inputs:} \texttt{data}, \texttt{treatment}, \texttt{period}, \texttt{type} (\texttt{"simple"}|\texttt{"complex"})\\
\textbf{Output:} carry-over dummy variables (\texttt{matrix}/\texttt{data.frame})};

\node (gee) [processGEE, below=2.8cm of carry]
{\textbf{\texttt{CrossGEE}} — standard GEE\\
\textbf{Inputs:} \texttt{id}, \texttt{carry}: from the createCarry output, \texttt{covariates} (opt.), \texttt{correlation} (\texttt{exchangeable},\texttt{AR(1)}, \texttt{unstructured}), \texttt{Mv} (opt.)\\
\textbf{Outputs:} estimates, variances, QIC};

\node (kron) [processKron, left=0.3cm of gee]
{\textbf{\texttt{CrossGEEKron}} — Kronecker correlation\\
\textbf{Inputs:} \texttt{time} (measurement occasions), plus all \texttt{CrossGEE} inputs\\
\textbf{Outputs:} estimates, variances, QIC};

\node (esp) [processESP, right=0.3cm of gee]
{\textbf{\texttt{CrossGEESP}} — semiparametric GEE\\
\textbf{Inputs:} \texttt{nodes} (opt., \# spline nodes), plus all \texttt{CrossGEE} inputs\\
\textbf{Outputs:} estimates, variances, QIC, effect plots};

\node (output) [startstop, below=0.6cm of gee]
{\textbf{Output}: estimated effects, QIC, and effect plots};

\draw[arrow] (start) -- (carry);
\draw[arrow] (carry) -- (gee);
\draw[arrow] (carry.east) to[out=0, in=90] (esp.north);
\draw[arrow] (carry.west) to[out=180, in=90] (kron.north);
\draw[arrow] (gee) -- (output);
\draw[arrow] (kron.south) |- (output.west);
\draw[arrow] (esp.south) |- (output.east);

\node (constraint) [note, below=0.5cm of carry, text width=8.8cm]
{\textbf{Carry-over constraint:} the \texttt{"complex"} carry-over option is \emph{only supported by \texttt{CrossGEESP}}. 
Use \texttt{"simple"} carry-over with \texttt{CrossGEE} and \texttt{CrossGEEKron}.};

\node[draw=gray!60, rounded corners, fit=(gee)(kron)(esp), inner sep=8pt,
      label={[gray!60]above:\footnotesize\textbf{GEE-based estimation}}] (geeGroup) {};

\node[draw=gray!60, rounded corners, fit=(start)(carry), inner sep=8pt,
      label={[gray!60]above:\footnotesize\textbf{carry-over encoding}}] (carryEnco) {};
      
\end{tikzpicture}
\caption{Flowchart of the \texttt{CrossCarry} package: from input to output, showing how \texttt{createCarry} feeds the three GEE variants (\texttt{CrossGEE}, \texttt{CrossGEEKron}, \texttt{CrossGEESP}) and their key inputs/outputs.}
\label{fig:CrossCarryFlowStyled}
\end{figure}
\begin{table}[ht]
\centering
\caption{Summary of the main functions in the \texttt{CrossCarry} package from an API perspective, including input types and constraints.}
\label{tab:CrossCarryAPI}
\begin{tabular}{|p{2.5cm}|p{7.5cm}|p{3cm}|}
\hline
Function & Key Inputs & Outputs \\
\hline
\texttt{createCarry} & 
\texttt{data} (data frame), 
\texttt{treatment} (factor/character), 
\texttt{period} (integer/numeric), 
\texttt{type} (character: ``simple" or ``complex") & 
Carry-over dummy variables (matrix/data frame) \\
\hline
\texttt{CrossGEE} & 
All \texttt{CreateCarry} inputs +
\texttt{id} (factor/character), 
\texttt{carry} (matrix/data frame), 
\texttt{covariates} (optional, matrix/data frame), 
\texttt{correlation} (character), 
\texttt{Mv} (optional, positive integer) & 
Parameter estimates, variance, QIC\\
\hline
\texttt{CrossGEEKron} & 
All \texttt{CrossGEE} inputs + 
\texttt{time} (numeric vector indicating measurement occasions) & 
Parameter estimates, variance, QIC \\
\hline
\texttt{CrossGEESP} & 
All \texttt{CrossGEE} inputs + 
\texttt{nodes} (optional, integer: number of spline nodes) & 
Parameter estimates, variance, QIC, effect plots \\
\hline
\end{tabular}
\end{table}
To provide a clearer overview of the package functions, Table~\ref{tab:CrossCarryAPI} summarizes each main function in terms of its key inputs and outputs, complementing Figure~\ref{fig:CrossCarryFlowStyled} and helping users understand the structure and usage of the package.  Complete manual description is located in C8 of Table \ref{codeMetadata}. The four main functions were selected to reflect the core steps in crossover-data analysis: carry-over encoding, marginal modelling, Kronecker-structured repeated measures, and semiparametric extensions {\citep{ken15}}.

\begin{itemize}
    \item \texttt{createCarry}: This function constructs a set of dummy variables indicating the presence of first-order carry-over effects. The user can specify whether the carry-over effect is \emph{simple}, or \emph{complex}. {Complex carry-over effects arise when the residual impact of a treatment in a given period depends on the combination of the previous and current treatments. Unlike a simple carry-over effect, which assumes the prior treatment affects all subsequent periods equally, the complex effect models treatment-sequence-specific residuals, capturing order-dependent interactions between consecutive treatments.}

    \item \texttt{CrossGEE}: This function performs generalized estimating equation (GEE) analysis of crossover designs assuming that the response variable belongs to the exponential family, following \cite{cruz2023analysis}. The correlation structure of the repeated measures can be specified as \emph{independence}, \emph{exchangeable}, \emph{AR1}, or \emph{unstructured}. Model selection criteria such as {Quasi-likelihood under the Independence model Criterion ($QIC$), its unbiased version ($QICu$), and the Correlation Information Criterion ($CIC$)} are provided to evaluate goodness-of-fit and select the optimal correlation structure and covariates. 

    \item \texttt{CrossGEEKron}: Extends \texttt{CrossGEE} by allowing a Kronecker product correlation structure to model repeated measures within and between periods. This function is particularly suited for designs where correlations within periods are assumed homogeneous and correlations between periods are proportional to the within-period correlation, following \cite{cruz2023correlation}.

    \item \texttt{CrossGEESP}: Implements semiparametric estimation of both parametric and nonparametric effects using penalized splines, including time effects and carry-over effects. It extends the \texttt{CrossGEEKron} methodology to allow for flexible modeling of complex carry-over effects, as detailed in \cite{cruz2023semiparametric}. The number of spline nodes can be specified by the \texttt{nodes} parameter, and the output includes model estimates, QIC values, and graphs of estimated effects. The selection of spline knots and penalty parameters follows the methodology detailed in \cite{cruz2025penalizedgeecomplexcarryover}.
\end{itemize}

Building on established GEE methodology, the package is designed to support robust estimation and inference for both simple and complex carry-over effects. It can be applied to a broad range of crossover experimental designs. {In previous methodological work, we conducted a series of simulation studies evaluating the estimation procedures implemented in the package. The first study \cite{cruz2023analysis} (function \texttt{CrossGEE}) showed that classical GEE estimators in two‐period crossover designs remained unbiased and retained correct type I error under moderate correlation misspecification. A second study \cite{cruz2023semiparametric} (function \texttt{CrossGEESP}) examined semiparametric extensions with heterogeneous carry-over and nonlinear time trends, demonstrating robustness to missing data, short washout periods, and complex temporal patterns, with notable efficiency gains over standard GEE. The third study \cite{cruz2023correlation} (function \texttt{CrossGEEKron}) evaluated Kronecker‐structured correlation models and found consistently lower RMSE and higher power than alternative working correlations, with QIC reliably identifying the true structure. Finally, \cite{cruz2025penalizedgeecomplexcarryover} (extended \texttt{CrossGEESP}) assessed penalized-spline estimators, showing near‐nominal coverage, improved RMSE, and stable reconstruction of smooth components. Collectively, these studies provide theoretical and comprehensive empirical validation for the methods implemented in the current software.}

Further technical details and theoretical properties of the estimators implemented in the package are documented in \cite{cruz2023semiparametric,cruz2023analysis,cruz2023correlation}.

\section{Illustrative examples}

\subsection{Data description}

The \texttt{Arterial} data set \cite{ken15} contains measurements of blood pressure in a crossover experimental design with three periods and three treatments. Each subject was measured multiple times within each period, and the variables are: \textbf{Pressure}: the response variable (blood pressure), \textbf{Subject}: unique identifier for each experimental unit, \textbf{Treatment}: the treatment applied in each period,  \textbf{Period}: the period of observation, and \textbf{Time}: the minute of measurement within each period.

\subsection{Example of Implementation}
\begin{figure}[ht]
\centering
\includegraphics[width=13cm]{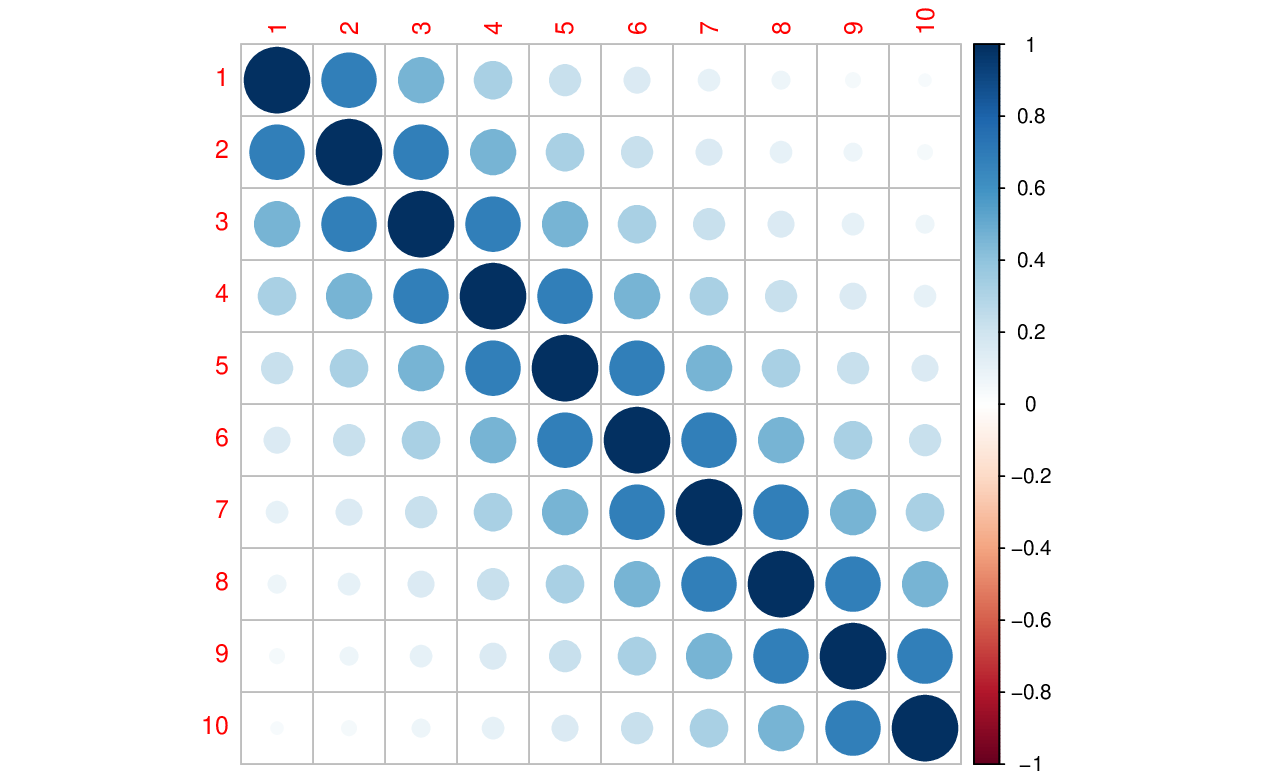}
\caption{Within-period correlation matrix for Arterial data obtained from \texttt{CrossGEEKron}.}
\label{wit}
\end{figure}

The following code demonstrates how to generate carry-over variables and fit different GEE models using this data:
\begin{lstlisting}[language=R, caption={create Carry-over dummy variables on the \texttt{Arterial} data}]
library(CrossCarry)
data("Arterial")
carrydata <- createCarry(data = Arterial, treatment = "Treatment", period = "Period", id = "Subject", carrySimple = TRUE)
data <- carrydata$data
carry <- carrydata$carryover
\end{lstlisting}
When the option \texttt{carrySimple=TRUE} is specified, the function generates two dummy variables corresponding to the simple carry-over effects. In particular, the variable \texttt{Carry\_B} takes the value 1 if the experimental unit received treatment B in the immediately preceding period, and 0 otherwise. In this setting, the functions \texttt{CrossGEE}, \texttt{CrossGEEKron}, and \texttt{CrossGEESP} fit a model including Treatment, Carry-over, Period, and Time as covariates. We fitted four GEE models with different correlation structures (CrossGEE, CrossGEEKron, and CrossGEESP). For brevity, only the code of CrossGEEKron is shown below, while the complete analysis, including all models, is available in the reproducible capsule (C3) in Table \ref{codeMetadata}.
\begin{lstlisting}[language=R, caption={Fitting GEE model on the \texttt{Arterial} data}]
model3 <- CrossGEEKron(response = "Pressure", treatment = "Treatment", period = "Period", id = "Subject", time = "Time",  carry = carry, data = data, correlation = "AR-M", Mv = 1)
\end{lstlisting}
\begin{table}[ht]
\centering
\begin{tabular}{lcccc}
\hline
 & Model 1 & Model 2 & Model 3 & Model 4 \\
\hline
QIC        & 49211.541 & 49007.588 & 48935.859 & 56046.283 \\
QICu       & 49172.494 & 48980.341 & 48910.690 & 55999.728 \\
Quasi Lik  & -24579.247 & -24483.170 & -24447.345 & -27994.864 \\
CIC        & 26.524 & 20.624 & 20.585 & 28.277 \\
Parameters & 7 & 7 & 8 & 5 \\
QICC       & 49259.541 & 49055.588 & 49025.859 & 56049.179 \\
\hline
\end{tabular}
\caption{Comparison of GEE models fitted to the Arterial data using the QIC criterion.}
\label{tab:QIC}
\end{table}
Table~\ref{tab:QIC} summarizes the QIC values across four fitted models. Based on this criterion, Model 3 provides the best fit.
The output of the \texttt{QIC} is displayed, which allows comparison of different models fitted to the same variable. Based on this criterion, Model 3 is selected as it provides the best fit among the three models considered. The following function then presents the summary of Model 3.
\begin{lstlisting}[language=R, caption={Summary of GEE model on the \texttt{Arterial} data}]
> summary(model3$model)
Coefficients:
            Estim  Naive SE Naive z Robust SE Robust z
(Intercept) 108.8787   2.75   39.61      3.1779    34.26
Period2       4.2239   3.50    1.21      2.1938     1.93
Period3       3.1747   3.22    0.99      1.2546     2.53
TreatmentB    0.8616   2.81    0.31      1.5590     0.55
TreatmentC   -5.8117   2.81   -2.07      1.5060    -3.86
Carry_B      -4.8511   3.60   -1.35      2.3319    -2.08
Carry_C      -5.8315   3.60   -1.62      2.4063    -2.42
Time         -0.0053   0.01   -0.52      0.0048    -1.11
\end{lstlisting}
This procedure provides two correlation matrices: the within-period correlation matrix $\pmb{R}_1(\pmb{\alpha}_1)$, and the between-period correlation matrix $\pmb{R}_2(\pmb{\alpha}_2)$ defined in \cite{cruz2023correlation}. The within-period correlation matrix $\pmb{R}_1(\pmb{\alpha}_1)$ is visualized in Figure \ref{wit}, showing an autoregressive decay of correlation across measurements.

\section{Impact}
The \texttt{CrossCarry} package has significantly advanced the analysis of crossover designs with repeated measures by providing tools that were previously unavailable or insufficient in standard R packages. 

It allows users to estimate both simple and complex carry-over effects using semiparametric methods \citep{cruz2023semiparametric}, and to model intra- and inter-period correlations through specialized GEE correlation structures \citep{cruz2023correlation}. 
These capabilities enable new research questions related to treatment effects under complex experimental designs, such as the impact of varying carry-over structures on longitudinal outcomes, which could not be reliably investigated before. 
Several studies have applied the \texttt{CrossCarry} methodology to real-world experiments, demonstrating its practical relevance and growing adoption. Selected examples are listed below, while the complete collection of publications is available in the Supplementary Material.
\begin{itemize}
    \item \cite{hurzeler2025cannabidiol} -planned and analyzed clinical trials on cannabidiol effects.
    \item \cite{ginting2024effects} - analyzed educational interventions using crossover designs.
    \item \cite{corea2025effects} - analyzed effects of substitution of corn with napier on dairy cows.
\end{itemize}
{In our experience supporting applied projects, CrossCarry has simplified the creation of carry-over effect matrices, automated the estimation of complex models, and made it easier for practitioners to obtain inference for non-normal responses in crossover designs.} By offering a unified framework for both standard and repeated-measures crossover designs, the package reduces methodological barriers, allowing researchers to focus on substantive questions rather than on cumbersome statistical programming. 

Currently, the package continues to evolve {and contributions from the community are welcome. A “how to contribute” guideline has been added to the GitHub repository, providing instructions for reporting issues, submitting feature requests, or proposing improvements via pull requests.} We are developing penalized GEE methodologies to further extend the functionality of \texttt{CrossCarry} beyond B-splines, and allowing regularization to improve estimation stability in high-dimensional settings. 
This ongoing development demonstrates the package's commitment to providing cutting-edge tools for the analysis of crossover designs and its potential to support a wide range of future research applications. This evidence demonstrates that \texttt{CrossCarry} has enabled new research questions, improved the analysis of existing crossover studies, and is actively being used by researchers internationally. The upcoming extension to penalized generalized estimating equations will broaden its applicability and allow more complex designs to be analyzed effectively.

{While \texttt{CrossCarry} provides a user-friendly interface for implementing advanced methods in crossover designs, users are cautioned that it does not replace the need to understand the underlying statistical principles. Familiarity with concepts such as marginal modelling, carry-over effects, and correlation structures is essential to correctly interpret results and avoid misapplication. For this reason, the package is supported by four methodological papers \cite{cruz2023analysis, cruz2023correlation, cruz2023semiparametric, cruz2025penalizedgeecomplexcarryover}, which provide detailed descriptions of the formulation, advantages, and limitations of the implemented analyses.}
\section{Conclusions}
The \texttt{CrossCarry} package facilitates the analysis of crossover designs with multiple configurations, under the assumption that the response variable belongs to the exponential family. For designs with only one observation per period, the package assumes simple carry-over effects, allowing straightforward analysis even in settings with limited data.

The incorporation of the Kronecker correlation structure enables precise estimation when multiple observations per individual are collected within each period. This structure accounts for both within-period and between-period correlations, improving the accuracy and reducing bias in treatment effect estimates.

Semiparametric GEE models in the package provide a flexible and robust approach for crossover designs with five or more observations per period. They allow modeling of complex carry-over effects while accommodating correlation structures typical in repeated measures designs.

Overall, \texttt{CrossCarry} enhances analytical capabilities for crossover studies by combining simple and complex carry-over modeling, semiparametric estimation, and flexible correlation structures. It provides a reliable and efficient tool for researchers seeking accurate inference in crossover designs across diverse experimental settings.

\end{document}